\documentclass[twocolumn, times]{aastex62}
\usepackage{natbib}
\usepackage{newtxmath}
\usepackage{color}
\hypersetup{linkcolor=blue,citecolor=blue,filecolor=blue,urlcolor=blue}

\newcommand{\pc}{\mathrm{pc}}

\newcommand{\Msol}{\textup{M}_\mathrm{\sun}}
\newcommand{\xMsol}[2]{\ensuremath{{#1}\times 10^{#2} \,\Msol}}

\newcommand{\Mvir}{\mathrm{M}_{200}}

\newcommand{\rvir}{\mathrm{r}_{200}}
\newcommand{\rhalflight}{\mathrm{r}_{1 /2}}
\newcommand{\iron}{\big[ \mathrm{Fe}\, / \mathrm{H} \big]}
\newcommand{\magv}{\mathcal{M}_V}
\newcommand{\vdisp}{\sigma_{\star}}

\newcommand{\MR}{\color{black}}

\graphicspath{{./}{figs/}}

\shorttitle{Genetically modified ultra-faints}
\shortauthors{Rey et al.}

\begin{document}

\title{EDGE: The origin of scatter in ultra-faint dwarf stellar masses and surface brightnesses
}



\correspondingauthor{Martin Rey}
\email{martin.rey.16@ucl.ac.uk}


\author[0000-0002-1515-995X]{Martin P. Rey}
\affil{Department of Physics and Astronomy,
University College London, London
WC1E 6BT, United Kingdom}

\author[0000-0001-9546-3849]{Andrew Pontzen}
\affil{Department of Physics and Astronomy,
University College London, London
WC1E 6BT, United Kingdom}

\author{Oscar Agertz}
\affil{Lund Observatory, Department of Astronomy and Theoretical Physics, Lund University, Box 43, SE-221 00 Lund, Sweden}

\author{Matthew D. A. Orkney}
\affil{Department of Physics, University of Surrey, Guildford, GU2 7XH, United Kingdom}

\author[0000-0002-1164-9302]{Justin I. Read}
\affil{Department of Physics, University of Surrey, Guildford, GU2 7XH, United Kingdom}

\author[0000-0003-4357-3450]{\quad \quad \quad Am\'elie Saintonge}
\affil{Department of Physics and Astronomy,
University College London, London
WC1E 6BT, United Kingdom}

\author[0000-0002-4315-9295]{Christian Pedersen}
\affil{Department of Physics and Astronomy,
University College London, London
WC1E 6BT, United Kingdom}

\begin{abstract}

We demonstrate how the {\MR least luminous} galaxies in the universe, ultra-faint dwarf galaxies, are sensitive to their dynamical mass at the time of cosmic reionization. We select a low-mass ($\sim \xMsol{1.5}{9}$) dark matter halo from a cosmological volume, and perform zoom hydrodynamical simulations with multiple alternative histories using ``genetically modified" initial conditions. Earlier-forming ultra-faints have higher stellar mass today, due to a longer period of star formation before their quenching by reionization. Our histories all converge to the same final dynamical mass, demonstrating the existence of extended scatter ($\geq$ 1 dex) in stellar masses at fixed halo mass due to the diversity of possible histories. One of our variants builds less than 2 \% of its final dynamical mass before reionization, rapidly quenching in-situ star formation. The bulk of its final stellar mass is later grown by dry mergers, depositing stars in the galaxy's outskirts and hence expanding its effective radius. This mechanism constitutes a new formation scenario for highly diffuse ($\rhalflight \sim 820 \, \pc$, $\MR \sim 32 \, \text{mag arcsec}^{-2}$), metal-poor ($\iron = -2.9$), ultra-faint ($\magv = -5.7$) dwarf galaxies within the reach of next-generation low surface brightness surveys.
\end{abstract}


\section{Introduction} \label{sec:intro}

The advent of digital, wide sky photometric surveys is revealing a vast population of low surface brightness galaxies. At the faintest end with $V$-band magnitudes {\MR $\magv \geq -8$}, are ``ultra-faint" dwarf galaxies, which are amongst the lowest-mass objects able to form stars in a $\Lambda$CDM universe (see \citealt{Simon2019} for a review).

Analysis of stellar populations within ultra-faint dwarf galaxies reveals that their stars have typical ages approaching that of the Universe (e.g. \citealt{Brown2014, Weisz2014}). This implies an early truncation of star formation, thought to arise because the galaxies' potential wells are too shallow to accrete and cool gas once cosmic reionization has heated the surrounding intergalactic medium at $z\sim6$ (\citealt{Efstathiou1992}). Reionization is powered by the entire population of galaxies and quasars, and therefore, to a first approximation, can be modelled without taking account of local conditions (though see \citealt{Katz2019}). However, galaxies with a given halo mass today have formed at different rates over cosmic time, and therefore had a wide range of masses at the time of reionization. This may lead to a large diversity in the properties of ultra-faint galaxies, {\MR depending on the specific coupling between the galaxy's history in a chosen cosmological model and the timing of cosmic reionization (e.g. \citealt{BenitezLlambay2015, Sawala2016, Fitts2017, Bozek2019}).}

Quantifying this expected scatter will be key to interpreting findings from forthcoming surveys, e.g., the Large Synoptic Survey Telescope (LSST). In particular, the low-mass of ultra-faints makes them particularly suited to identifying any fingerprints of alternative dark matter models (see \citealt{Pontzen2014} for a review). Meeting this promise requires us to model the formation of galaxies with a range of cosmological histories, each with sufficient resolution to resolve the interstellar medium and astrophysical processes within such small objects {\MR (\citealt{Maccio2017, Wheeler2019, Agertz2019,Munshi2019})}. 

{\MR In this second paper of the Engineering Dwarfs at Galaxy Formation's Edge (EDGE) project, we couple cosmological high-resolution zoom simulations (\citealt{Agertz2019}) to the genetic modification framework (\citealt{Roth2016, Pontzen2017, Rey2018}).} This method generates alternative initial conditions for a cosmological galaxy, each new version varying a specific aspect of the galaxy's mass accretion history (hereafter MAH). Each history is simulated independently, reproducing the same large-scale environment and final dynamical mass. This enables a controlled study, allowing us to construct a causal account of the link between history and observables.

\section{Genetically modified dwarf galaxies} \label{sec:setup}

We first create zoom initial conditions (see \citealt{Agertz2019} for a detailed description of the procedure) for an unmodified, reference galaxy with a present-day virial mass of $\Mvir = \xMsol{1.5}{9}$, where $\Mvir$ defines the mass enclosed within a sphere of radius $\rvir$ encompassing $200$ times the critical density of the universe. This galaxy is chosen as an isolated central, with no massive neighbors within $5 \, \rvir$. 

We construct ``genetically modified,'' alternative initial conditions for this galaxy, modifying its halo mass around reionization, while fixing the halo mass today. The halo mass of an object can be directly controlled from the initial conditions by modifying the height of its associated density peak (\citealt{Roth2016}). We therefore identify all particles within the major progenitor at redshift $z=6$ and increase (decrease) the mean density within this region to increase (decrease) the halo mass at $z=6$. We conserve the final, $z=0$, halo mass by maintaining a constant mean density in the corresponding Lagrangian region. We emphasize that each modified initial condition makes minimal changes to the surrounding environment, maintaining the same large-scale filamentary structure around the galaxy (e.g. Figure 1 of \citealt{Rey2019}).

We evolve the modified and reference initial conditions to $z=0$ using cosmological zoomed galaxy formation simulations. We follow the evolution of dark matter, stars, and gas using the adaptative mesh refinement hydrodynamics code \textsc{ramses} (\citealt{Teyssier2002}). We include an extensive galaxy formation model described in detail by \citet{Agertz2019} as ``Fiducial." {\MR We model cosmic reionization as a time-dependent ultraviolet (UV) background}{\footnote{We use an updated version of \citet{Haardt1996} as implemented in the public \textsc{ramses} distribution.}} {\MR and account for gas cooling and heating. We model star formation by stochastically forming star particles in gas cells with densities above $300 \, m_{\text{H}} \, \text{cm}^{-3}$ (\citealt{Agertz2019}) and stellar feedback using energy, mass, and metallicity budgets described in \citet{Agertz2013}.} A key feature of our simulations is their resolution, greatly reducing uncertainties in modelling {\MR the injection of feedback from supernovae}. We refine down to a maximum spatial resolution of $3 \, \pc$ and follow dark matter particles with masses $960 \, \Msol$. This resolution is sufficient to capture the cooling radius of individual supernovae, allowing us to directly inject thermal energy in a gas cell and self-consistently follow the buildup of momentum by solving the hydrodynamics equations (\citealt{Kim2015, Martizzi2015}).

\begin{figure}
  \centering
    \includegraphics[width=\columnwidth]{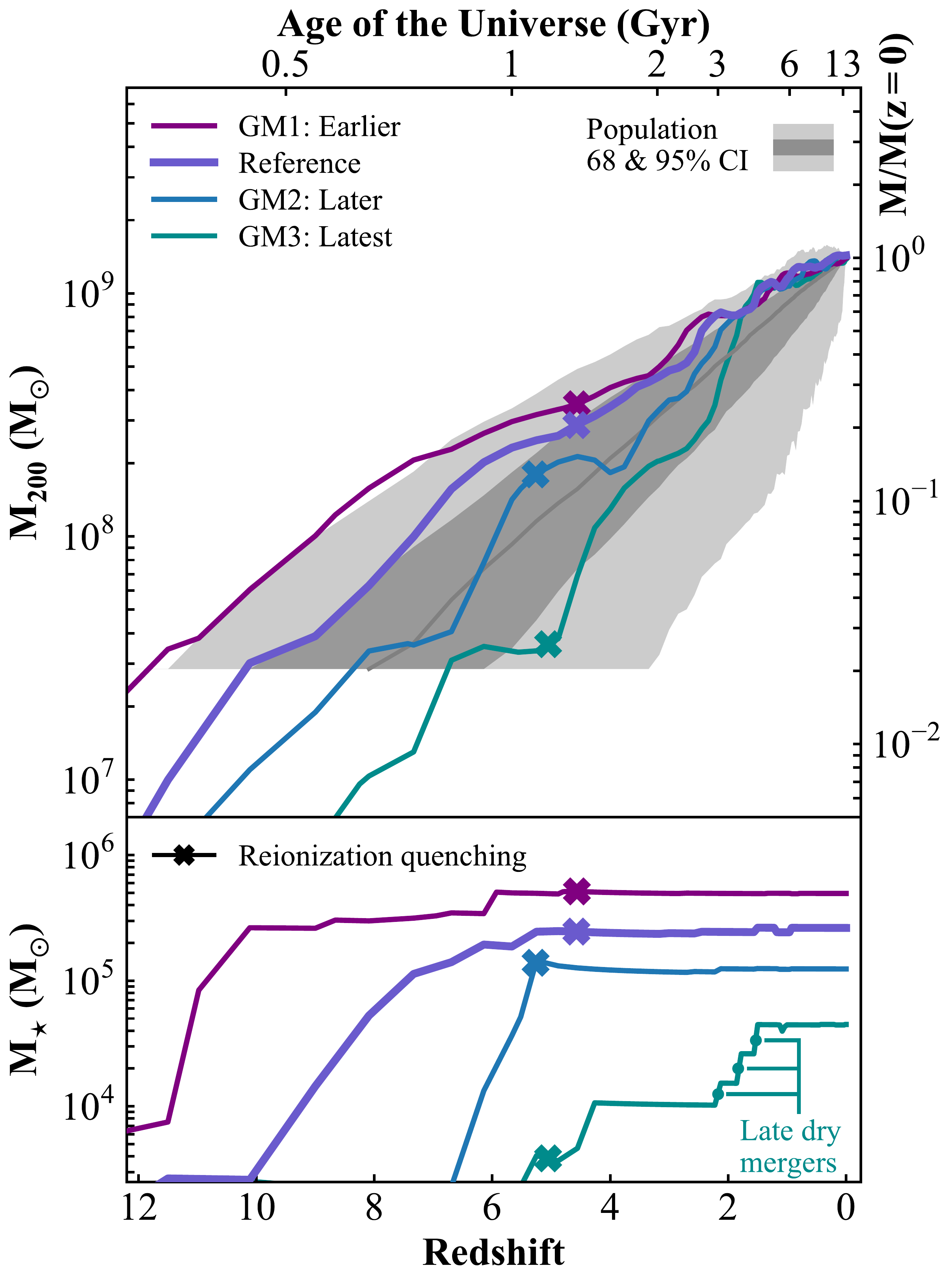}

    \caption{Growth of dynamical mass (top panel) over cosmic time for our reference galaxy and its three genetically modified counterparts. We design each new history to vary the formation time of the reference galaxy, forming earlier (purple) or later (blue, turquoise). Earlier-forming ultra-faints start assembling stellar mass (bottom panel) earlier in time, before reionization quenches in-situ star formation by preventing gas inflows (crosses). Earlier-forming galaxies therefore have a systematically higher $z=0$ stellar mass. By construction, all histories converge to the same dynamical mass today while scanning across a representative range of
    early histories (grey bands). This allows us to quantify scatter in the stellar mass at fixed halo mass (see Figure~\ref{fig:mstarmhalo}).}

    \label{fig:massgrowth}

\end{figure}

Despite the accurate modelling of supernovae explosions, additional feedback channels can strengthen or weaken their coupling to the surrounding interstellar medium (e.g. \citealt{Agertz2019, Smith2019}). To probe the sensitivity of our results to such residual uncertainties, we evolve all initial conditions with an alternative model, reducing the efficiency of supernova feedback (``Weak feedback" model in \citealt{Agertz2019}). {\MR This model introduces arbitrary temperature and velocity ceilings for supernovae ejecta ($10^8 \, \text{K}$ and $1000 \, \text{km} \, \text{s}^{-1}$ respectively), thereby limiting their efficiency in driving winds and regulating star formation. We stress that this model should be seen as an explorative test, rather than as an alternate physical prescription.}

We identify halos using the \textsc{hop} halo finder (\citealt{Eisentein1998}) and construct their mass histories using the \textsc{pynbody} \citep{Pontzen2013} and \textsc{tangos} \citep{Pontzen2018} libraries. For each galaxy, we compute the $V$-band luminosity of each stellar particle using a single stellar population model interpolated over a grid of ages and metallicities (\citealt{Girardi2010}) and sum them to derive the total magnitude. We choose a random line-of-sight to obtain the projected half-light radius, $\rhalflight$, and checked that using unprojected 3D half-light radius does not modify our observed trends. Finally, we compute the {\MR one-dimensional stellar velocity dispersion as $\vdisp = \sqrt{\sigma_{\star, \text{x}}^2 + \sigma_{\star, \text{y}}^2 + \sigma_{\star, \text{z}}^2} \, / \sqrt{3}$ and the total iron metallicity, $\iron$, as the mean of each stellar particle's iron abundance weighted by its stellar mass (\citealt{Escala2018, Agertz2019})}.

\section{Growing the stellar mass of ultra-faints} \label{sec:mstar}

We show the resulting four genetically modified MAHs in Figure~\ref{fig:massgrowth}, top panel. Our modifications generate a range of halo mass growth before cosmic reionization, while these histories have converged by $z \sim 2$ and reach the same dynamical mass today. To illustrate that we probe a cosmologically representative range of histories, we compare these tracks with a statistical sample of $\sim 1500$ histories extracted from the parent, lower-resolution volume of our zoom simulations. We select central halos with masses {\MR as defined by the halo finder} between $0.9$ and $\xMsol{4}{9}$ at $z=0$ and compute their fractional mass growth, i.e., their mass growth divided by their total mass. We show the median with 64 \% and 95 \% confidence intervals at each redshift (grey bands), normalised to $\xMsol{1.5}{9}$. The four MAHs lie within the 95\% contours of the overall population, demonstrating that our objects range across the majority of the population's scatter in early histories. We stress that this comparison should be seen as qualitative rather than as a rigorous statistical test; we leave more detailed statistical inference to a future work.

In the bottom panel of Figure~\ref{fig:massgrowth}, we show the growth of stellar mass of each genetically modified history. As an ultra-faint dwarf galaxy forms earlier (i.e., achieves a higher mass at reionization) its final stellar mass grows. Since the environment surroundings and final mass are all fixed, this trend demonstrates a direct mapping between the halo mass achieved before reionization and the final stellar mass of an ultra-faint.

The link is best explained by the duration of the star-forming phase for each galaxy. Figure~\ref{fig:massgrowth} shows that earlier-forming galaxies systematically start building stellar mass earlier in time. We mark by a cross the time of the last star formation activity in the main progenitor, showing that reionization quenches star formation at near-identical times ($z \sim 4$) for all MAHs. Star formation continues shortly after the end of reionization from self-shielded cold gas within the halo (\citealt{Onorbe2015}). However, heating from the UV background prevents further gas accretion, quickly leading to starvation and permanent quenching. Earlier-forming ultra-faints therefore have a longer period of star formation. In addition, earlier-forming ultra-faints have a higher halo mass at a given time and consequently reach higher instantaneous star formation rates before reionization. 

\begin{figure}
  \centering
    \includegraphics[width=\columnwidth]{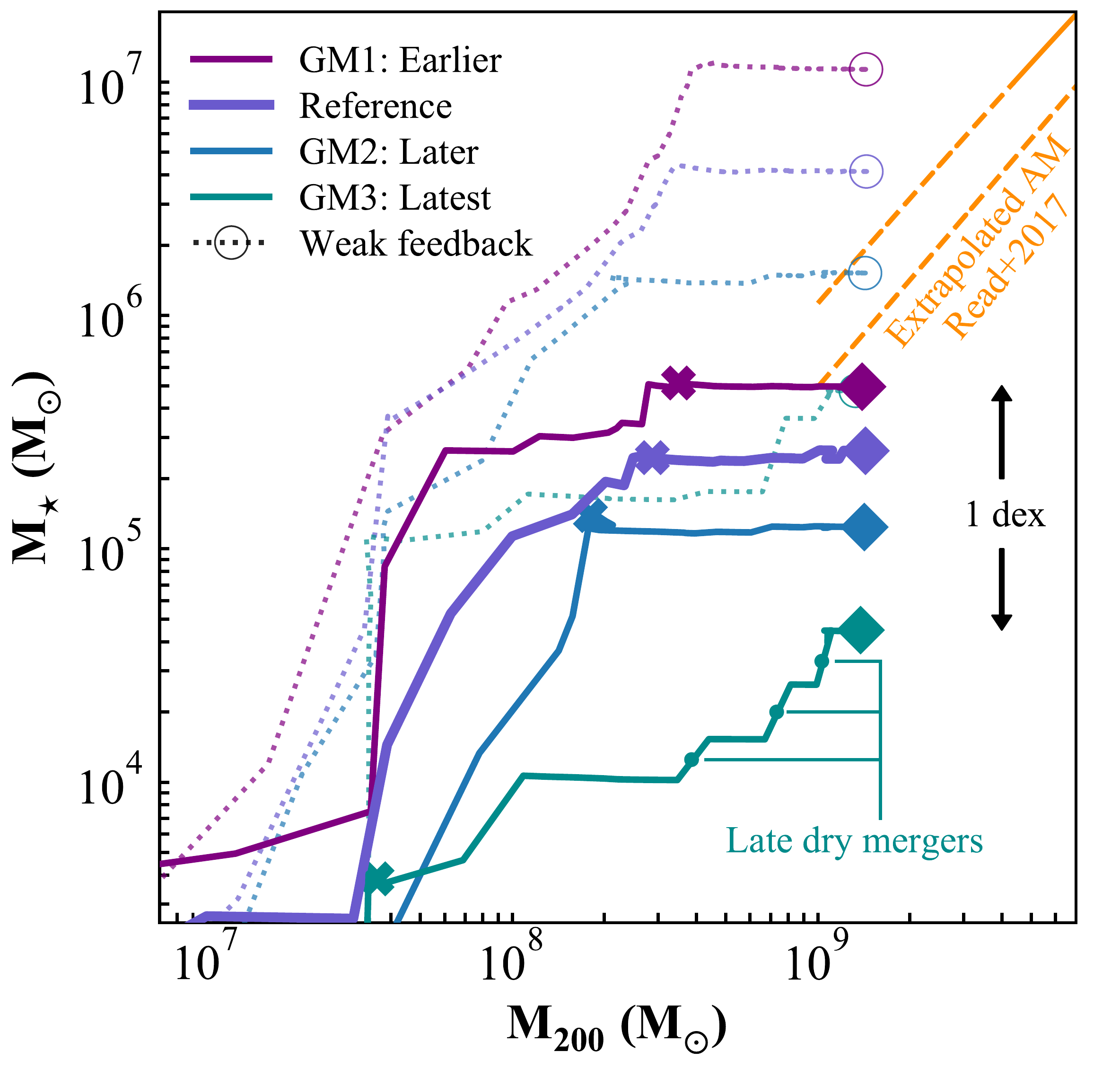}

    \caption{Stellar mass and halo mass growth of our four modified galaxies compared. The range of histories creates a 1 dex spread in stellar masses at fixed halo mass. Varying the implementation of supernova feedback (faint dotted) modifies the overall normalization of the stellar masses, but relative differences between histories are conserved. This robustly demonstrates the existence of extended scatter in the relation between stellar mass and halo mass due to the variety of possible formation times. Our latest-forming history (turquoise) further sees its mode of stellar mass growth modified, now being dominated by late-time dry stellar accretion. This new formation scenario has a strong impact on observable structural properties (Figure~\ref{fig:rstar}).}
    \label{fig:mstarmhalo}
\end{figure}

More extended and more vigorous star formation thus leads to a higher stellar mass today. We now examine the consequences for the relation between stellar mass and halo mass.

\section{Generating scatter in the stellar mass-halo mass relation} \label{sec:mhalomstar}

The mass of a galaxy's dark matter halo is thought to be the primary driver of its properties, as it regulates the depth of the potential well and hence the  overall availability of gas. This assumption allows empirical and semi-numerical models of galaxy formation to rely on a parameterized mapping between halo mass and stellar mass (see \citealt{Wechsler2018} for a review). On the scale of dwarf galaxies, however, the functional form of this mapping is highly uncertain (e.g. \citealt{Behroozi2013, Brook2014, GarrisonKimmel2017, Read2017}). Our results above further show how the assumption of a one-to-one correspondence between stellar mass and halo mass breaks down at this mass scale.

We show in Figure~\ref{fig:mstarmhalo} the growth of our modified galaxies in the stellar mass-halo mass plane. The final $z=0$ points (color diamonds) can be compared with an extrapolated abundance matching prediction (orange dashed, \citealt{Read2017}). Our different histories generate a spread in stellar mass over 1~dex, causally demonstrating the existence of extended scatter due to the variety of possible histories for a given ultra-faint. 

Since our galaxies have by construction the same environment to isolate the role of histories, the extent of the exposed scatter is a lower bound on the overall population diversity. External factors such as tidal stripping during the infall into a more massive host can provide an additional mechanism to generate scatter on the scale of ultra-faints (\citealt{Munshi2017}).

To probe the robustness of our results to residual uncertainties in modelling galaxy formation, we show in Figure~\ref{fig:mstarmhalo} our four modified histories evolved with the alternative, ``Weak feedback," model. As expected, all histories have higher stellar masses compared to the ``Fiducial" model (solid lines). However, the systematic trend of higher stellar mass with earlier-forming galaxies remains, while the scatter in stellar masses at fixed halo mass increases from 1 to 1.3 dex. This approximate conservation of scatter reflects its origin in the relative timing of mass accretion and reionization (Section 3). We conclude that the strength of supernova coupling primarily affects the \textit{absolute} scale of stellar masses, while the diversity in formation times drives the relative scatter around the mean. 

In addition to varying the total stellar mass, our modifications change
the mode of stellar mass growth for the latest-forming MAH (turquoise). Rather than forming stars in-situ, this galaxy accretes most of its stellar mass through late, dry mergers (labeled in Figures~\ref{fig:massgrowth} and~\ref{fig:mstarmhalo}). We now show that this accretion-dominated scenario exposes a new formation pathway for extended, diffuse ultra-faint dwarf galaxies.

\begin{figure}
  \centering
    \includegraphics[width=\columnwidth]{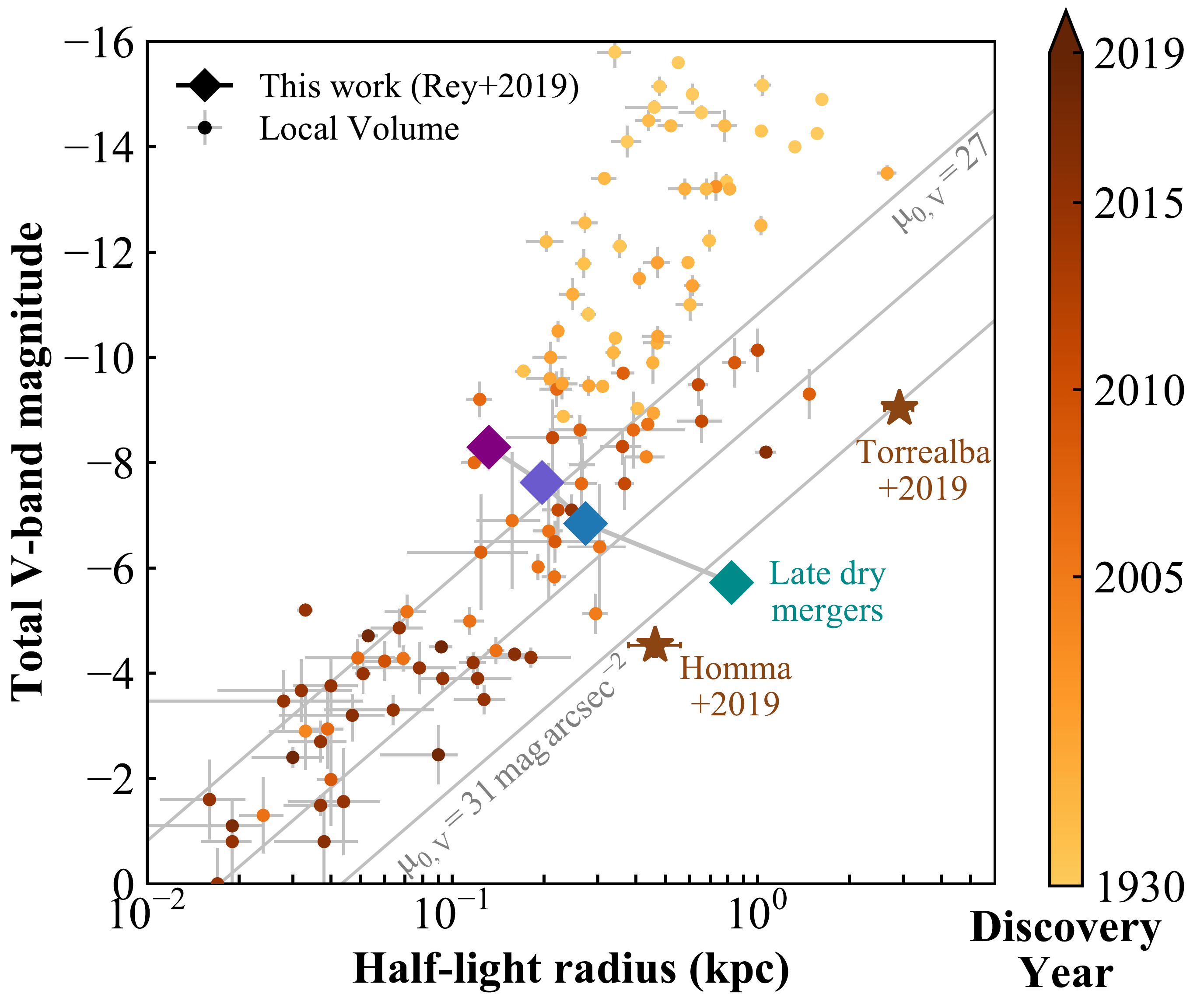}

    \caption{Impact of modifying the history of an ultra-faint on its $V$-band magnitude and projected half-light radius. We compare our galaxies to a sample of observed dwarfs from the Local Volume (\citealt{McConnachie2012, Kirby2014, Simon2019}). All our simulated galaxies lie within current observational scatter except the latest-forming dwarf that builds its stellar mass through dry mergers. These mergers deposit stars on wide orbits, creating an extremely diffuse ultra-faint. We color individual galaxies by their discovery year, highlighting two trends: toward fainter detections overall and more extended galaxies at a given magnitude. Our dwarf forming through dry mergers has a central surface brightness comparable to the latest discoveries of modern deep imaging surveys (e.g. \citealt{Homma2019, Torrealba2019}), highlighting prospects to uncover this diffuse population with, e.g. LSST and future GAIA releases.}
    \label{fig:rstar}

\end{figure}

\section{The formation of a diffuse ultra-faint}

Figure~\ref{fig:rstar} plots the total $V$-band magnitude at $z=0$ against the projected half-light radius of our genetically modified galaxies (color diamonds). A compiled sample of observations from the Local Volume (\citealt{McConnachie2012, Kirby2013, Kirby2014, Simon2019}) is shown for comparison. We show lines of constant central surface brightness (\citealt{Martin2016}, Eq 8), assuming a mean ellipticity of 0.5 (\citealt{Simon2019}). The three earliest-forming galaxies (purple, violet, blue) all lie within the observational scatter, while the latest-forming (turquoise) has a lower surface brightness than currently detected dwarf galaxies. 

The extreme diffuse nature of this object arises due to its assembly from ex-situ accreted stars. Dry stellar mergers after $z=6$ build $94\%$ of the total stellar mass of this galaxy, depositing stars away from the galaxy centre and growing the galaxy's half-light radius to $820 \, \pc$ today. The result is an extremely extended ultra-faint with low central surface brightness ($\MR \sim 32 \, \text{mag arcsec}^{-2}$). We therefore have exposed a new formation scenario for extended diffuse ultra-faint dwarf galaxies, arising through an early truncation of in-situ star formation by reionization but a later growth via ex-situ dry accretion. The mass growth associated to this object is within the 68\% contour of the overall population (see Figure~\ref{fig:massgrowth}), likely making this scenario generic for field ultra-faint dwarfs. 

The current rarity of observed analogs for our diffuse ultra-faint reflects observational challenges at this extremely low surface brightness. Each observed dwarf galaxy in Figure~\ref{fig:rstar} is colored by the year of their discovery. Two trends are visible as experiments improved over the years: (i) downward toward overall fainter objects and (ii) rightward toward more diffuse objects at a given magnitude. Ongoing and next-generation surveys are expecting to pursue these trends, vastly expanding the census of ultra-faint dwarf galaxies. In particular, detections through resolved stars are reaching surface brightnesses similar to our diffuse galaxy, as demonstrated by two recent candidates for Milky-Way satellites (\citealt{Homma2019, Torrealba2019}). LSST will further be able to detect individual stars of a $\mathcal{M}_V = -6$ galaxy out to several Mpc, directly probing ultra-faint formation in the field. {\MR We therefore expect our prediction of the existence of ultra-faint, diffuse field galaxies to be testable in the near future. 

Future data releases from GAIA will also continue to extend the Milky-Way census toward lower surface brightnesses. Our results apply to isolated ultra-faints and thus cannot be readily compared with the Milky-Way satellite population. They nonetheless provide a base for studying the fate of such low surface brightness objects as they fall into a massive host that we will study in future work.} 

Spectroscopic follow-up of stars within such a diffuse galaxy could confirm its formation pathway. Its stellar velocity dispersion $\vdisp = 7.4 \, \text{km s}^{-1}$ is comparable to that of our reference galaxy ($7.1 \, \text{km s}^{-1}$), highlighting the weak constraining power of stellar dispersions on formation scenarios. However, the assembly from multiple low stellar mass systems produces an extremely metal-poor galaxy with $\iron = -2.9$, compared to $-2.4$ in our reference case where 88\% of the total stellar mass is formed in-situ. A low metallicity at a given stellar mass would therefore complement the extended size as a signature of this new formation scenario.

\section{Conclusion}

We have presented results demonstrating how formation history affects the properties of field ultra-faint dwarf galaxies. We created a series of four ``genetically modified" zoom initial conditions (\citealt{Roth2016, Rey2019}) for a single object, systematically varying its accretion history up to the time of cosmic reionization while fixing its $z=0$ dynamical mass. We evolved these initial conditions with high-resolution zoom cosmological simulations (\citealt{Agertz2019}) and computed the response of the central galaxy's observables.

By construction, all our histories converge to the same dynamical mass today and evolve within the same cosmological environment, thereby creating a controlled study. We demonstrate that the halo mass achieved before reionization directly controls the final stellar mass of an ultra-faint (Figure~\ref{fig:massgrowth}). Earlier-forming galaxies begin forming stars when the universe was younger and have a more vigorous star formation rate at a given time, therefore assembling higher stellar masses before their quenching by reionization. We further show that the variety of possible histories for an ultra-faint leads to an extended scatter in the relation between stellar mass and halo mass (Figure~\ref{fig:mstarmhalo}). This scatter arising from histories is robust to a large variation in our implementation of stellar feedback.

Probing the interaction between merger histories and reionization allows us to expose the potential for highly diffuse, ultra-faint dwarf galaxies {\MR in the field}. Extremely low surface brightness can be achieved through an early truncation of in-situ star formation and a later growth by stellar accretion, vastly growing stellar size (Figure~\ref{fig:rstar}). Finding such a population will be within the reach of future facilities such as LSST.

Our study cleanly demonstrates the importance of cosmological histories in explaining the diversity of dwarf galaxy properties. This extends previous results (e.g. \citealt{Fitts2017}) by (i) targeting smaller dynamical mass halos more likely to host observed ultra-faints (\citealt{Jethwa2018, Read2019AM}) (ii) improved numerical resolution, reaching the critical scale of the supernova cooling radius, and (iii) cleanly isolating the role of histories using the genetic modification technique. Nonetheless, the interaction between histories and reionization is only one factor in determining the full diversity of the ultra-faint dwarf galaxy population (\citealt{Sawala2016}). In future work, we will investigate other sources of diversity such as environment (e.g. \citealt{Munshi2017}) and halo mass in a larger suite of objects (Orkney et al. 2019 in preparation).

\acknowledgments

{\MR We thank Vasily Belokurov, Alyson Brooks, Marla Geha, Risa Wechsler, and the anonymous referee for comments that improved the clarity of this manuscript.} M.R. acknowledges support from the Perren Fund and the IMPACT fund. A.P. and A.S. are supported by the Royal Society. O.A. acknowledges support from the Swedish Research Council (grant 2014-5791) and the Knut and Alice Wallenberg Foundation. M.O. would like to thank the STFC for support (grant ST/R505134/1). This project has received funding from the European Union’s Horizon 2020 research and innovation programme under grant agreement No. 818085 GMGalaxies. This work was partially supported by the UCL Cosmoparticle Initiative. The authors acknowledge the use of the UCL Grace High Performance Computing Facility, the Surrey Eureka supercomputer facility and associated support services. This work was performed in part using the DiRAC Data Intensive service at Leicester, operated by the University of Leicester IT Services, which forms part of the STFC DiRAC HPC Facility (www.dirac.ac.uk). 

\bibliography{Biblio}
\end{document}